# Hypervisor Extension for a RISC-V Processor

J. Gauchola[1], J. Costa[1], E. Morancho[1], R. Canal[1], X. Carril[2], M. Doblas[2], B. Otero[1], M. A. Pajuelo[1], E. Rodríguez[1], J. Salamero[2] and J. Verdú[1]

[1]Computer Architecture Department, Universitat Politècnica de Catalunya  [2]Barcelona Supercomputing Centre

**Abstract**

*This paper describes our experience implementing a Hypervisor extension for a 64-bit RISC-V processor. We describe the design process and the main required parts with a brief explanation of each one.*

## Introduction

Virtual environments are widely used to share the resources of a physical machine between several users or processes, such as Virtual Machines (VMs). In other words, a single physical machine can execute multiple VMs concurrently. E.g., most data centers and cloud services use VMs to isolate the users between them.

Another utility of VMs is the ability to execute several Operating Systems (OSs) concurrently on the same physical machine, besides the native one. These OS can be the same than the native one or a completely different one, e.g. executing Linux in a virtual environment meanwhile the native OS is Windows.

These VMs are managed by a Virtual Machine Monitor which controls the execution and behavior of the VMs. This control can be implemented by software, but it adds a significant performance overhead. Therefore, hardware support, like the Hypervisor extension in RISC-V specification[1], can mitigate this overhead.

In the scope of the Designing RISC-V-based Accelerators for next generation Computers (DRAC) project[1], we have designed and implemented the Hypervisor extension for a RISC-V processor, the Lagarto core [2], able to run Linux OS but without supporting the Hypervisor extension.[2]

## Background

We will implement the Hypervisor extension on a RISC-V core that follows the Sv39 specification, where given a 64-bit virtual address just the 39 lower bits are significant. This restriction impacts on the size of the Page Table Entries (PTEs) of the page tables.

The core has three privilege modes: M (Machine), S (Supervisor) and U (User). A similar work but on a different core has been detailed in [3].

[1] https://drac.bsc.es/
[2] This work is partially supported by the DRAC (IU16-011591), the HORIZON Vitamin-V (101093062) and the Computación de Altas Prestaciones VIII (PID2019-107255GB) projects.

## Design

Adding the Hypervisor extension to an existing RISC-V core requires adding new privilege modes and modifying the following modules: Control and Status Registers (CSR), Trap Handling, and the Address Translation (specially, the Page Table Walker and the Translation Lookaside Buffer).

### New virtual privilege modes

When the Hypervisor extension is enabled, two new privilege modes arise: VU (Virtual User) and VS (Virtual Supervisor). VS is devoted to execute the guest OS and VU to user processes in the guest OS. Also, S mode becomes HS (Hypervisor-Extended Supervisor).

### Control and Status Registers (CSRs)

RISC-V specification defines a 4096-register address space that, according to the current privilege mode, keeps configuration and status information of the core (exception codes, root page table, performance counters, ...). Each one of these entries is a CSR.

The hypervisor extension specifies the behaviour of multiple CSRs under the new privilege modes, adding a new set of registers and modifying the existing ones. We have designed the update procedure of these CSRs when the relevant events take place.

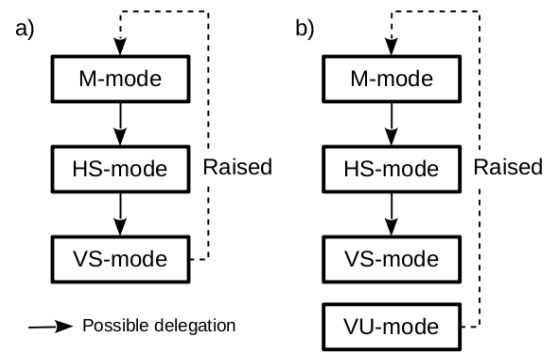

**Figure 1:** *Trap delegation: a) from VS, b) from VU*

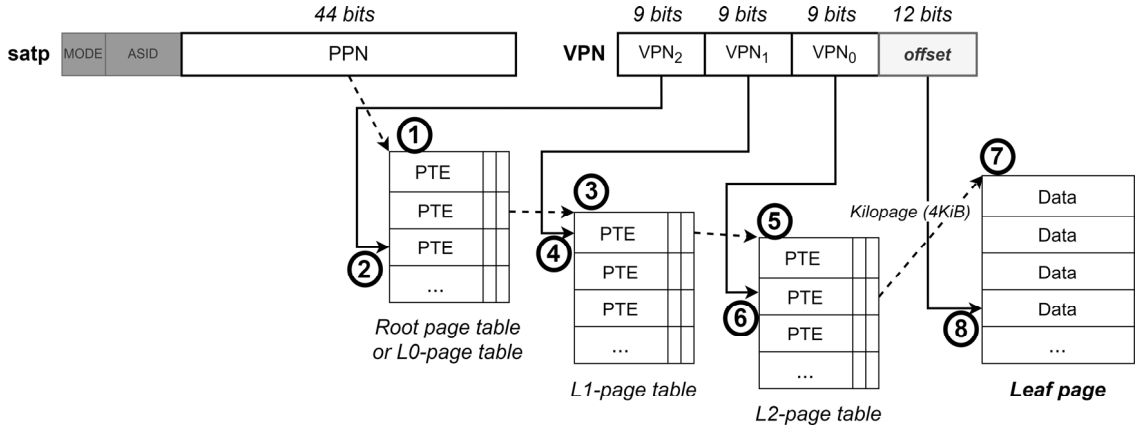

**Figure 2:** *Sv39 address translation process where a Virtual address (VPN + offset) is translated to its Physical address(⑧) by using a 3-level page table (①– ⑦).*

## Trap Handling

Adding new privilege modes has a great impact on the trap handling system. According to the privilege mode of the core when the trap is raised and to some CSRs, each trap is delegated to a handling routine running in a specific privilege mode. Figure 1 shows the possible delegations of traps raised in the new modes.

## Address Translation

Address translation in RISC-V relies on memory paging. Sv39 supports three page sizes (4 KiB, 2 MiB and 1 GiB) and the page table is organized as an up-to 3-level tree of page tables. The root page table is pointed by the satp CSR.

Conceptually, all address translations require traversing this tree structure (Figure 2). The Page Table Walker (PTW) is responsible for this traversing. The resulting translation is stored in the Translation Look-aside Buffer (TLB) to enable faster translations in the future.

When the virtualization is enabled, execution mode of the core is virtualized (e.g. VU) and the address translation becomes a two-stage procedure (Figure 3). Firstly, addresses generated in VU-mode (*Guest Virtual Addresses*, GVAs) are translated into *Guest Physical Addresses* (GPAs) by the guest OS and the VM. Secondly, the GPAs are translated finally into *Host or Supervisor Physical Addresses* (HPA or SPA) by the native OS and the physical machine.

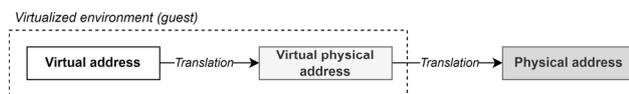

**Figure 3:** *Two-stage address translation*

The Hypervisor extension has a significant impact both on TLB and PTW to deal with the two-stage address translation. The problem is that all the physical addresses on a guest virtual machine must be translated, and this means not just the resulting physical address from the guest, but also the physical addresses of the page tables used by the translation itself.

## Testing

New RISC-V assembler programs were made to check the functioning of the hypervisor extension, such as the new hypervisor CSRs, new trap handling and different two-stage address translation levels.

## Conclusions

This paper presents the modifications required to add an initial implementation of the Hypervisor extension on the Lagarto core. This modifications can be summarized as adding support to:

- New privilege levels (VU, VS, HS).
- Trap handling mechanism.
- Two-stage virtual memory address translation.

Adding this support enables the use of this processor in data centers and cloud environments where efficient virtualization is a requirement.